 \documentclass[smallabstract,smallcaptions]{dccpaper}

\usepackage{epsfig}
\usepackage{citesort}
\usepackage{amsmath}
\usepackage{amssymb}
\usepackage{color}
\usepackage{url}
\usepackage{svg}
\usepackage{listings}
\usepackage{graphicx}
\newlength{\figurewidth}
\newlength{\smallfigurewidth}

\begin{document}

\title
{\large
\textbf{Entropy Coding Improvement for Low-complexity Compressive Auto-encoders}
}

\author{%
Franck Galpin, Muhammet Balcilar, 
Frederic Lefebvre, Fabien Racapé,   Pierre Hellier \\[0.5em]
{\small\begin{minipage}{\linewidth}\begin{center}
\begin{tabular}{c}
InterDigital, Inc.  \\
Rennes, France \\
\url{firstname.lastname@interdigital.com} 
\end{tabular}
\end{center}\end{minipage}}
}

\maketitle
\thispagestyle{empty}

\begin{abstract}
End-to-end image and video compression using auto-encoders (AE) offers new appealing perspectives in terms of rate-distortion gains and applications. While most complex models are on par with the latest compression standard like VVC/H.266 
on objective metrics, practical implementation and complexity remain strong issues for real-world applications. In this paper, we propose a practical implementation suitable for realistic applications, leading to a low-complexity model. We demonstrate that some gains can be achieved on top of a state-of-the-art low-complexity AE, even when using simpler implementation. Improvements include off-training entropy coding improvement and encoder side Rate Distortion Optimized Quantization. Results show a $19\%$ improvement in BDrate on basic implementation of fully-factorized model, and $15.3\%$ improvement compared to the original implementation. The proposed implementation also allows a direct integration of such approaches on a variety of platforms.
\end{abstract}

\section{Introduction}

Rate-distortion autoencoders \cite{habibian2019video} form the backbone of modern neural codecs, where the latent representation is conditioned to optimize a rate-distortion loss function. These methods are a special case of Variational Autoencoder (VAE) \cite{kingma2013auto} with three key differences: (i) the posterior distribution is a uniform distribution centered on the encoder's outputs (latents) at training time, (ii) has a fixed variance output distribution and (iii) has trainable priors \cite{theis2017lossy,balle2016end}. It was shown that minimizing the evidence lower bound (ELBO) of this special VAE is equivalent to minimizing jointly the mean square error (MSE) of the reconstruction and the entropy of latents w.r.t the priors \cite{balle2018variational}. Existing models differ mainly by the modelling of the priors and selection of encoder/decoder architecture. The simplest model used a small number of convolutional layers and learned priors by non-parametric way called fully-factorized model \cite{balle2016end}. This model was then improved by the hyperprior model, where the parametric prior is a function of side information and modelled by zero-mean gaussian \cite{balle2018variational}, gaussian \cite{minen_joint,minnen2020channel} or a mixture of gaussian \cite{cheng2020image}. These models are deeper and use an encoder-decoder pair for the side information. From an architectural perspective,  attention layers \cite{cheng2020image,gao2021neural}, frequency decomposition layers \cite{gao2021neural} and invertible layers \cite{xie2021enhanced} are recent developments to improve the codec's performance. All those codecs with improved performance on top of the original design increase the complexity and the practicability of the  decoder for two main reasons. Firstly, the complexity (measured for example in MAC/sample) is increased because of the added networks and layers. Secondly, the parallelism potential is reduced and the latency increased because of the addition of the hyper-prior and the use of an autoregressive schema for context adaptive coding.

The most performing neural codecs are almost on par with the latest compression standard on generic image compression and propose many advantages over standard codecs, such as easy adaptation on perceptual distortion metrics and high performance on specific domains thanks to their learning ability. However, they are currently impractical on small devices because of the computational complexity and energy consumption. It has been reported that a neural decoder needs more than $1000$ times the number of MACs/pixel compared to standard codecs \cite{le2022mobilecodec} \cite{Bossen2021}. Even with network quantization and distillation like in \cite{van2021instance}, the necessary number of MACs can be around $500$ times more compared to standard codecs. These complexity issue is one of the most important problem that makes neural codec not practical for many applications.  

Since learning-based models are trained on large datasets, they may be optimal in average for all train set, but they are likely to be suboptimal for any single image, where this problem is called the amortization gap \cite{cremer2018inference}. In image compression context, this problem can be solved by improving rate-distortion objective at encoding time for a given single image. In literature, the proposed methods to reduce this gap are three folds. The first ones involved latent finetuning for the given image  \cite{Aytekin_2018_CVPR_Workshops,LuCZCOXG20,Campos_2019_CVPR_Workshops,NEURIPS2020_066f182b,Guo_2020_CVPR_Workshops}, the second ones model parameters finetuning  \cite{9088301,Lam_Yat_Hong,9287069,van2021overfitting,biastune} for better performance on a given single image or the re-parameterization of the entropy model in order to better fit the latents without fine-tuning the model \cite{reducinggap}. 
In practice, only the first class of method allows to keep the same decoder and does not increase the decoder complexity.

In this paper, we propose a practical implementation of a low-complexity AE.
More specifically, we propose to demonstrate a practical implementation on a simple a model, derived from a fully-factorized model \cite{balle2016end} using the entropy parametrization found in \cite{balle2018variational} as implemented in 
in \cite{compressai}:
\begin{itemize}
    \item we replace the greedy gdn/igdn activations by Relu/Relu activations to obtain a more hardware friendly implementation. 
    \item we distillate the neural codec's with a full $16$-bits integer network using simplified integerized operations using the SADL framework \cite{sadl}.
    \item we learn a new context switching based conditional entropy model on top of the latent representation, initially learned with factorized entropy model. 
    \item we propose a new rate distortion optimization process at encoding time in order to decrease the amortization gap, without the need of an external framework, using only the distillated decoder.
\end{itemize}
 


We first present the general process to train a compressive auto-encoders. We then present the process to derive an integerized version of the model. A conditional entropy model is then computed from the extracted model. The latent optimization stage is then presented. Results on the Kodak dataset are finally presented.




\section{Compressive Auto-encoders}
\label{sec:endtoend}
\subsection{Overview}
In this section, we introduce the basic principles of fully-factorized compressive auto-encoder including the training, inference and quantization steps. Let $\mathbf{x},\mathbf{\hat{x}}\in \mathbf{R}^{n\times n \times 3}$ be respectively an RGB image and its reconstruction, with a $n \times n$ size (the image is considered square without any loss of generality). Let $\mathbf{y},\mathbf{\hat{y}}\in \mathbf{R}^{m\times m \times s}$ be respectively the continuous latent and the quantized latent (or noise added), where $m \times m$ is the spatial resolution and $s$  is the number of channels. $\mathbf{Q}(.)$ is element-wise function that applies nearest integer quantization at test time (or its continuous relaxation at train time as $\mathbf{Q}(x)=x+\epsilon$ with $\epsilon \sim U(-0.5,0.5)$). 

Let us now summarize all the steps of the compressive autoencoder at test time. The sender inputs the image $\mathbf{x}$ to obtain the continuous latent $\mathbf{y}=g_a(\mathbf{x};\mathbf{\phi})$ and its version $\mathbf{\hat{y}}=\mathbf{Q}(\mathbf{y})$. The Quantized latents $\mathbf{\hat{y}}$ are then converted into a bitstream using the learned entropy model $p_{f}(.|\Psi)$. The receiver decodes the quantized latents $\mathbf{\hat{y}}$ from the bitstream using the shared entropy model $p_{f}(.|\Psi)$ and the reconstructed image is obtained as $\mathbf{\hat{x}}=g_s(\mathbf{\hat{y}};\mathbf{\theta})$. The encoder block $g_a$, decoder block $g_s$ and entropy model $p_{f}$ are trainable models implemented with neural networks, and parameterized by $\mathbf{\phi}$, $\mathbf{\theta}$ and $\Psi$ respectively.

The compressive auto-encoder optimizes $\mathbf{\phi}$, $\mathbf{\theta}$ and $\Psi$ by minimizing two objectives simultaneously. The first one is any differentiable distortion loss between $\mathbf{x}$ and $\mathbf{\hat{x}}$, while the second one is the length of the bitstream encoding $\mathbf{\hat{y}}$. Since $\mathbf{\hat{y}}$ is losslessly encoded using any entropy encoder such as range asymmetric numeral systems (RANS) \cite{duda2009asymmetric}, and because RANS is asymptotically optimal encoder, according to Shanon's entropy theorem the lower bound of bitlength can be used as an objective function. Compared to the experimental bitlength from RANS, that leads to a differentiable objective function. As a result, the loss function of the fully-factorized compressive auto-encoder can be written as follows;
\begin{equation}
   \label{eq:balle1}
   \mathcal{L}=\mathop{\mathbb{E}}_{\substack{\mathbf{x}\sim p_x \\ \epsilon \sim U}}\left[-log(p_{f}(\mathbf{\hat{y}}|\Psi)) + \lambda.d(\mathbf{x},\mathbf{\hat{x}})\right].
\end{equation}

Here, $d(.,.)$ is any distortion loss such as MSE for PSNR metric, $\lambda$ is a trade-off hyperparameter between compression ratio and quality, and $-log(p_{f}(\mathbf{\hat{y}}|\Psi))$ is the bitlength lower bound of encoded $\mathbf{\hat{y}}$. In order to entropy encode/decode the quantized latents, RANS needs probability mass function (PMF) of each quantized latent in $\mathbf{\hat{y}}$. the fully-factorized entropy model implements it by learning $s$ number of cumulative distribution function (CDF) shown by $\bar{p}_{\Psi}^{(c)}(.):\mathbf{R} \rightarrow \mathbf{R},
~for~ c=1\dots s$ implemented by neural networks. Under nearest integer quantization, the necessary PMFs can be derived from CDFs by $\hat{p}_{\Psi}^{(c)}(x)=\bar{p}_{\Psi}^{(c)}(x+0.5)-\bar{p}_{\Psi}^{(c)}(x-0.5)$. Since each PMF is dedicated for single $m \times m$ slice of latent (for each feature channel) and treat them independently, the entropy model applies as follows in \cite{balle2016end}:

\begin{equation}
   \label{eq:factorent}
   p_{f}(\mathbf{\hat{y}}| \Psi)=\prod_{c=1}^{s} \prod_{i,j=1}^{m,m} \hat{p}_{\Psi}^{(c)}({\mathbf{\hat{y}}_{i,j,c}}).
\end{equation}


\subsection{Architecture}
In our work we propose to improve over the baseline \cite{balle2016end} with the entropy bottleneck proposed in \cite{balle2018variational}, where 
\begin{itemize}
    \item $g_a$ employs $4$ convolutional layers (and subsample) and $3$ nonlinear Generalized Divisive Normalizations (GDN)
    \item $g_s$ employs $4$ deconvolutional layers (convolutional and upsample) and $3$ nonlinear inverse Generalized Divisive Normalizations (iGDN)
\end{itemize}
The GDN is a composition of linear transformations, followed by a generalized form of divisive normalization. This activation/normalization layer includes a division and square root, which is not suitable for practical hardware implementation. As in \cite{balle2018integer}, the GDN operation is replaced by ReLU, which is a more implementation friendly activation.


\subsection{Quantization}
To avoid using floating point operations when using the Neural Network at inference, the network uses only 16 bits integer arithmetic. It means that both network parameters and latent should be quantized and all operations are replaced by integer equivalent operations.

This proposal uses a static quantization approach. However, to further reduce the complexity, the integerized network use a simplified quantization approach (see \cite{sadl}:
\begin{itemize}
    \item the quantized parameters or latent do not use zero point, avoiding some additional operations,
    \item the quantizers are only power of 2, allowing to perform operations only using bit shifting (instead of costly multiplication/division),
\end{itemize}

\section{Entropy coding}

\subsection{Post-training conditional entropy}

\subsubsection{Context modeling}
Starting from a fully-factorized model with a basic entropy model, we enhance it by computing a conditional entropy model post-training. The method can be applied on any entropy model. We apply it here on a non parametric entropy model using Cumulative Distribution Functions (CDF).

To limit the latency and the computational complexity of the entropy decoder, a simple context modeling is performed to select the distribution used for decoding a latent value which can be deduced using additions and comparisons only.
In detail, for each value of the latent $v_{i,j,k}$ where $i,j$ are the spatial components and $k$ the channel component, its context is selected as
\begin{equation}\label{context}
C = \left\{
    \begin{aligned}
    0 & \textrm{, if } v_{i-1,j,k}<\epsilon_k \wedge  v_{i,j-1,k}<\epsilon_k \wedge  v_{i,j,k-1}<\epsilon_{k-1}\\
    1 & \textrm{, if } v_{i-1,j,k}>=\epsilon_k \oplus  v_{i,j-1,k}>=\epsilon_k \oplus  v_{i,j,k-1}>=\epsilon_{k-1}\\
    3 & \textrm{, if } v_{i-1,j,k}>=\epsilon_k \wedge  v_{i,j-1,k}>=\epsilon_k \wedge  v_{i,j,k-1}>=\epsilon_{k-1}\\
    2 & \textrm{, otherwise}
    \end{aligned}
\right.
\end{equation}
where $\epsilon_k$ is a threshold defined per channel. The context $C$ is then used to choose among the 4 CDFs for a particular channel $k$.

\subsubsection{Channel activation}
Another proposed improvement of the channels coding is the use of an activation bit for each channel. A channel is considered activated when at least one value is different from the most probable value of the channel (as deduced from the channel distribution).

In order to encode the activation bit of the channel, the probability of activation of a channel is computed over a large dataset. The activation bit is then entropy coded using this probability.

\subsubsection{CDF computation}
Statistics on latent variables are gathered over a large dataset to compute the CDFs for each channel.
In practice, the CDFs are extracted as the normalized, strictly monotonous, cumulative histograms for each channel and each context in $[0..3]$ as shown in listing \ref{algo1}.

\begin{lstlisting}[caption={Extraction of contextualized CDF},label={algo1}]
    for all latents L in dataset
     for all channel C in L
        for all value v in C
            ctx = compute C(v)
            H[ctx][C] += 1
        end
     end
    end
    for i in 0..3
      CDF[i]=normalize(H[i])
    end
\end{lstlisting}

\subsubsection{Channel ordering}
In the above context modeling, it is assumed that the conditional entropy between 2 consecutive channels is low. As the training stage does not enforce such constraint, as opposed to a fully auto-regressive model, we perform a reordering of the channels for encoding.

We first defined 2 entropy coding methods:
\begin{itemize}
    \item K1 uses a spatial only conditional entropy coding, using 3 CDFs per channel to encode each variable of the latent, conditioned by the values of its top and left neighbors.
    \item K2 uses the full conditional entropy coding, conditioned by the values of both the spatial neighbors and the collocated variable in the previous channel. 
\end{itemize}

First, the channel with the highest entropy is selected to be the first encoded. K1 coding is used to select the channel over a dataset of latents. 
The channel giving the best entropy decrease is then selected iteratively for each next channel: for each remaining channel to encode, a full encoding using the K2 model is used to compute the entropy of the channel. The entropy difference compared to the entropy using the K1 model is computed. The channel having the best entropy gain is selected as the next channel to encode. The process is repeated until all channels have been sorted.

\subsubsection{Thresholds computation}
The computation of the thresholds used in eq. \ref{context} are done per channel. For each channel independently, the best threshold, which minimizes the entropy of the channel using the model K1, is computed. Each threshold is then associated with the channel and used to deduce the spatial and inter-channel based contexts.

\subsection{Inference based RDOQ}
Finally, a Rate-Distortion Optimized Quantization is performed on the latent. In the proposed method, we do not rely on a backward pass of the decoder, which assumes a differentiable loss as in \cite{Campos_2019_CVPR_Workshops}. Instead we directly use the decoder inference to optimize the latent, like in traditional codecs.
The overall naive process is to modify each value of the latent and to test if the RD-cost $D+\lambda R$ decreases at each step.
The advantages of this approach are the following:
\begin{itemize}
    \item it only relies on the availability of the decoder inside the encoder,
    \item no gradient computations are needed, which would require the original floating point model,
    \item no differentiable loss, especially for the rate term, is required, which would not be possible when using simplified context switching entropy models.
\end{itemize}
On the other hand, this approach can be computationally expensive since it requires an iterative update of the variables of the latent.
In order to reduce the complexity and speed-up the process, the following improvements are added:
\begin{itemize}
    \item for each variable, only a limited variation of the value is considered, typically $[-1,1]$. Moreover, the variation is only tested for values that are not already optimally entropy coded, i.e., for values which are not already the most probable value in the distribution.
    \item as the modification of a particular latent value only have an impact inside the receptive field centered around this value, only a subset of the latent centered around the value to test is used in order to reduce the complexity. Moreover, the impact of the value modification on the border of the receptive field are negligible in the distortion change. In practice, using only a subset of the tensor with a size half of the receptive field centered on the value to test is enough.
    \item to speed-up the process on multi-core architectures, the latent optimization is done in parallel over several channels. As the latent updates may not be deterministic, the process is repeated several times until convergence. In practice, we found that 3 passes are enough to converge towards a minimum for the rate distortion cost.
\end{itemize}

\section{Results}

\subsection{Description}
The performance are assess on the Kodak dataset for 7 bitrates in the range $[0,1.5]$ bpp. It corresponds to the lambda used during the training in the following range: $\{0.0018, 0.0035, 0.0067, 0.02, 0.04, 0.08, 0.0130\}$.

The test conditions are presented in Table \ref{table:tests}. All AE used the CompressAI framework and training conditions, with $200$ epochs. The two proposed methods are implemented using the lightweight framework SADL \cite{sadl} as a pair of standalone encoder/decoder in pure C++. It should be pointed that no additional frameworks are necessary, only encoder and decoder and the CDFs for each channel is needed.

\begin{table}[h]
{\footnotesize
\begin{center}
\caption{\label{table:tests} Tests list.}
\begin{tabular}{|c|c|c|c|c|}
\hline
Name &  Activation (enc/dec)  & Entropy model & Latent Optimization \\
\hline
Balle2018 & GDN/iGDN \cite{balle2016end} & CDFs simple \cite{balle2018variational} & none \\
Base & ReLU/ReLU & CDFs simple \cite{balle2018variational} & none \\
LatentTune & ReLU/ReLU & CDFs simple \cite{balle2018variational} & optimized latent \cite{Campos_2019_CVPR_Workshops}\\
FillGap & ReLU/ReLU & Re-parameterized CDF \cite{reducinggap} &  none \\
LatentTune+FillGap & ReLU/ReLU & Re-parameterized CDF \cite{reducinggap} &  optimized latent \cite{Campos_2019_CVPR_Workshops} \\
Contexts (ours) & ReLU/ReLU & contexts switching & none \\
Contexts+RDOQ (ours) & ReLU/ReLU & contexts switching & inference based RDOQ \\
HM 16.22 \cite{hm} & traditional codec & & \\
VTM 11.0 \cite{vtm} & traditional codec & & \\
\hline
\end{tabular}
\end{center}
}
\end{table}
The last two anchors are traditional codecs used for reference, using default AI configuration for testing. Performances are evaluated by computing the BDrate gains over the base anchor.

\subsection{Performance}
\begin{figure}[h]
\begin{center}
\includegraphics[width=13cm]{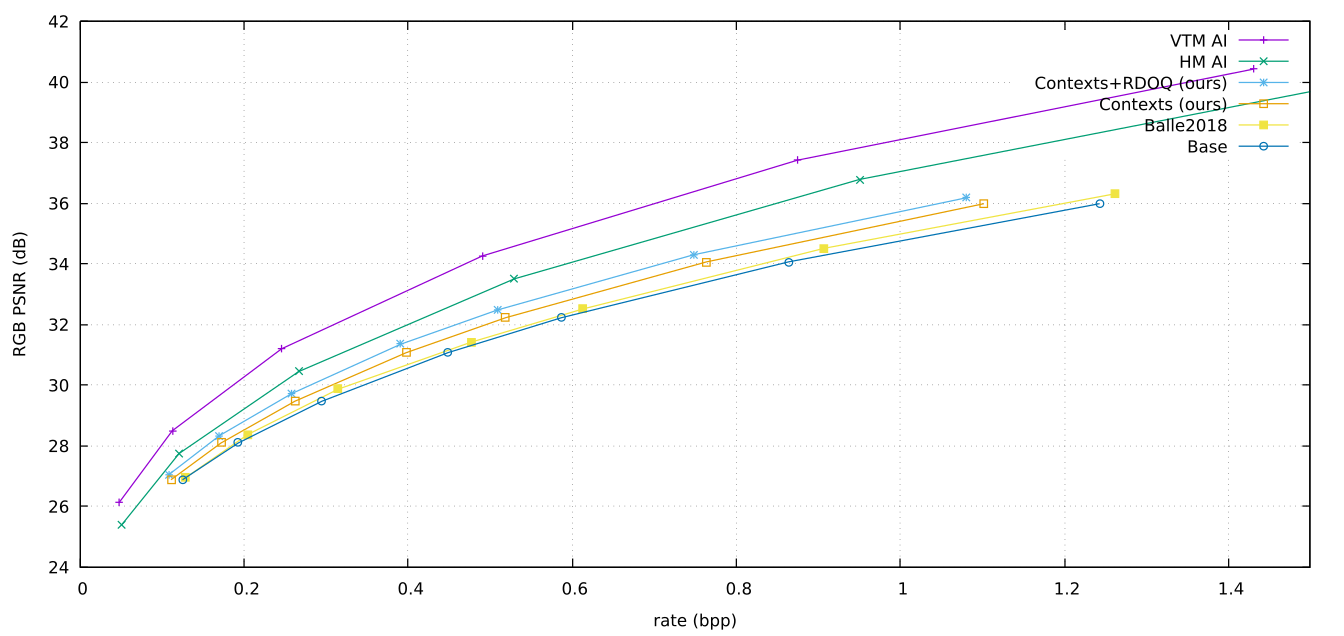}
\end{center}
\vspace*{-1cm}
\caption{\label{fig:perf}
RD curves of the different methods.}
\end{figure}

The figure \ref{fig:perf} shows the RD curves of the different methods over the bitrates range.
The results using the contexts switching entropy model and RDOQ allow to recover the loss from original Balle2018 model (using GDN, float model and pytorch implementation), having a simple model with about -15\% gains on top of this reference.
The table \ref{table:perf} shows the results in BDrate and also compared to similar methods using optimized entropy coding or latent optimization. 

\begin{table}[h]
\begin{center}
\caption{\label{table:perf}%
BDrate gains over base anchor.}
{
\footnotesize
\begin{tabular}{|c|r|}
\hline
Method & BDrate (\%) \\ 
\hline
Balle2018 & -3.15\% \\
FillGap &  -6.49\% \\
LatentTune & -10.86\% \\
Contexts & -11.07\% \\
LatentTune+FillGap & -15.09\% \\
Contexts+RDOQ & -18.56\% \\
HM 16.22 &  -30.93\% \\
VTM 11.0  & -46.99\% \\
\hline
\end{tabular}}
\end{center}
\end{table}


\subsection{Code}
All source code, training scripts, results and resulting codecs of the proposed methods are available at \cite{compressai}. The full codecs are standalone without any library dependencies, including the model inference. It allows easy integration in existing code. Moreover, complexity comparison with public implementation of traditional codecs which usually uses CPU, single threaded implementation (like the ones of HM or VTM) is easier.

\section{Conclusion}
We have proposed in this paper a practical implementation of a low complexity end-to-end auto-encoder. Even though our distillated decoder has some loss compared to the original base, the proposed encoder side optimizations and context based entropy coding allow to save $19\%$ bitrate compared to the distillated model, while saving $15.3\%$ compared to the original model. We believe these optimizations are the building bricks towards a neural codec that could be deployed.



\Section{References}
\bibliographystyle{IEEEbib}
\bibliography{refs}

\end{document}